\documentclass{ifacconf}

\usepackage{graphicx}      
\usepackage{natbib}        
\usepackage{xcolor}
\usepackage{amsmath,amssymb, amsfonts, mathtools}
\usepackage{enumerate}

\DeclareMathOperator*{\argmin}{arg\,min}

\newcommand{\mc}{\mathcal}

\DeclareMathOperator*{\subjto}{subject\,to}

\DeclareRobustCommand{\bigO}{%
  \text{\usefont{OMS}{cmsy}{m}{n}O}%
}

\begin{document}
\begin{frontmatter}

\title{A Receding Horizon Scheduling Approach for Search \& Rescue Scenarios\thanksref{footnoteinfo}} 

\thanks[footnoteinfo]{This work was supported by Siemens Corporate Technology.}

\author[First]{Yousef Emam} 
\author[First]{Sean Wilson} 
\author[Second]{Mathias Hakenberg} 
\author[Second]{Ulrich Munz}
\author[First]{Magnus Egerstedt}

\address[First]{Institute for Robotics and Intelligent Machines, Georgia Institute of Technology, Atlanta, GA 30332, USA (e-mail: \{emamy, sean.t.wilson, magnus\}@gatech.edu).}
\address[Second]{Siemens Corpororate Technology, Princeton, NJ 08540, USA (e-mail: \{mathias.hakenberg, ulrich.muenz\}@siemens.com)} 

\begin{abstract}                
Many applications involving complex multi-task problems such as disaster relief, logistics and manufacturing necessitate the deployment and coordination of heterogeneous multi-agent systems due to the sheer number of tasks that must be executed simultaneously. A fundamental requirement for the successful coordination of such systems is leveraging the specialization of each agent within the team. This work presents a Receding Horizon Planning (RHP) framework aimed at scheduling tasks for heterogeneous multi-agent teams in a robust manner. In order to allow for the modular addition and removal of different types of agents to the team, the proposed framework accounts for the capabilities that each agent exhibits (e.g. quadrotors are agile and agnostic to rough terrain but are not suited to transport heavy payloads). An instantiation of the proposed RHP is developed and tested for a search and rescue scenario. Moreover, we present an abstracted search and rescue simulation environment, where a heterogeneous team of agents is deployed to simultaneously explore the environment, find and rescue trapped victims, and extinguish spreading fires as quickly as possible. We validate the effectiveness of our approach through extensive simulations comparing the presented framework with various planning horizons to a greedy task allocation scheme.
\end{abstract}


\end{frontmatter}

\section{Introduction}
\label{sec:intro}

Multi-robot systems are well suited to solve complex tasks in dynamic and dangerous environments due to their redundancy, ability to operate in parallel, and system level fault tolerance to individual failure as highlighted in \cite{brambilla2013swarm, sahin2004swarm}. Multi-Robot Task Allocation (MRTA) deals with the assignment of agents to tasks in order to achieve an overall system goal within the constraints of the deployment setting. Therefore, in order to leverage the potential multi-robot systems have to successfully operate in dynamic and dangerous environments to solve complex problems such as disaster response, search and rescue, environmental monitoring, and automated warehousing, effective methods for solving the MRTA problem are needed (\cite{gerkey2004formal}).

The MRTA problem becomes more complex when introducing morphological or behavioral heterogeneity within a deployed multi-robot system (\cite{swarmanoid2013}). However, this additional complexity comes with the benefit of improving the overall system efficiency by leveraging the strengths of individual robots within the collective. For example, in a Search and Rescue (SaR) scenario, quadrotors, which are quick and agile, are better suited for scouting and surveying while ground robots are better suited for debris clearing and resource extraction. By leveraging these strengths efficiently and allocating tasks appropriately the heterogeneous system could out perform a system comprised of only aerial or ground robots.

 Moreover, in many scenarios the MRTA problem is accompanied by timing constraints where tasks must be performed sequentially, e.g. a robot must wait for a delivery before transporting the delivered package. Problems of this type are typically referred to as scheduling problems and involve an additional complexity. This class of problems can be solved with Mixed Integer Linear Programs (MILP) that attempt to schedule all the tasks at once, however, this approach suffers from an exponential complexity as noted in \cite{gombolay2013fast}. Additionally, when deploying a multi-agent system in dynamic environments, the system must be able to respond and reschedule tasks when unavoidable and inevitable environmental disturbances occur. Lastly and specific to the SaR operation following a natural disaster (e.g. wildfires, earthquakes, hurricanes) is that the team of agents must cover the targeted area and rescue victims amongst other tasks within a short-time window. This is due to the drastic decrease in the likelihood of victims surviving after $48$ hours as highlighted in \cite{48hours}. Therefore, the SaR problem can be cast as an instance of heterogeneous multi-agent system scheduling with the objective of minimizing the time of completion of all tasks (i.e. the makespan). 

Inspired by the work in \cite{rhp}, this paper proposes a Receding Horizon Planning framework to solve the heterogeneous MRTA scheduling problem and demonstrates its effectiveness in a SaR application. Inspired by Model Predictive Control, at fixed time intervals, the proposed framework detailed in Section~\ref{sec:rhp} schedules tasks for each agent up to a pre-defined time horizon and leverages a heuristic to estimate the cost to go for each schedule. As such, this framework does not suffer from the exponential complexity caused by the scheduling of the tasks and is robust to changes in the environment. Moreover, a specialized version of the framework for the SaR application is presented in Section~\ref{sec:useCase} along with a simulation environment for an abstracted SaR scenario. To validate the effectiveness of the proposed approach, Section~\ref{sec:experiments} presents extensive experimentation comparing the proposed Receding Horizon Planning based approach to a greedy scheduling scheme.

\section{Literature Review}
\label{sec:litReview}
As mentioned in Section~\ref{sec:intro}, the topic of coordination for multi-agent teams in SaR scenarios following natural disasters falls under the umbrella of Multi-Robot Task Allocation (MRTA). A comprehensive taxonomy of existing methods for MRTA can be found in \cite{gerkey2004formal}. As highlighted in \cite{khamis2015multi}, most existing approaches can be categorized as decentralized or centralized approaches. 

There are two main advantages to decentralized approaches: their robustness to varying team sizes and communication failures, and their scalability with respect to the size of the agent fleet thanks to the computational burden being shared amongst the agents. Moreover, market-based approaches such as \cite{coalitions,dias2006market,vig2006market}, attempt to combine the benefits of centralized and decentralized methods by having the computational burden shared between a central entity and the remainder of the fleet. For example, in \cite{coalitions}, the authors suggest a protocol where agents communicate their respective capabilities and use this information to form the coalitions in a decentralized fashion. As such, these methods are able to generate better solutions than fully decentralized approaches while maintaining a certain level of scalability. However, since SaR scenarios typically involve a bounded number of agents and little computational constraints, scalability with respect to the size of the team is of no concern; thus making fully-centralized approaches more suitable for this application.

Moreover, the topic of task allocation specifically pertaining to SaR scenarios is well-studied, most notably, by the participants of RoboCup SaR Agent Simulation competition as highlighted in \cite{Sheh2016}. The competition setup is as follows. A heterogeneous team is to be deployed to extinguish fires and rescue victims. Specifically, there are three types of agents: ambulances which rescue victims, fire brigades which extinguish fires and police units which remove the road blockades enabling the two other type of agents to reach their desired targets faster. The state-of-the-art task allocation strategy utilized by winning teams such as MRL in the competition is K-Means clustering of the Fires/Victims followed by a cluster to agent assignment using the Hungarian Algorithm which runs in polynomial time. We refer the reader to \cite{mrl19} for details.

Inspired by the RoboCup competition, the motivation behind the development of the new simulation environment presented in Section~\ref{sec:useCase} is two-fold. First, the proposed scenario can be seen as a generalized version of the competition's scenario. Specifically, instead of having a fixed number of types of agents each associated with a single class of tasks (e.g. police units only capable of removing road blockades), through characterizing each agent through the capabilities it exhibits, the proposed simulation framework allows for the modular addition of agent types and the collaboration of a heterogeneous sub-team of agents in achieving a single task. For example, given any two agents and their potentially different capacities to transport water, in the proposed scenario, they can indeed collaborate to extinguish a target fire. Moreover, since not all SaR scenarios are identical in nature, the proposed simulation environment frames the problem as a dynamic set of pick and place tasks where victims and resources are to be delivered to target locations.  

Based on this abstracted view of SaR problems, the Receding Horizon Planning framework presented in this paper aims to leverage the strengths of centralized scheduling approaches while keeping the problem size tractable and remaining robust to changes occurring in the environment. The robustness property is obtained through the repeated generation of schedules at fixed time intervals; whereas tractability of the problem size is obtained through only scheduling tasks up to a certain time horizon and leveraging a load-balancing Linear Program as a heuristic to estimate the cost-to-go. Consequently solutions produced by the proposed framework are not guaranteed global optimality since the schedules are of finite horizon. However, we present extensive empirical evidence demonstrating the effectiveness of the proposed approach in Section~\ref{sec:experiments}. In the next section, the proposed Receding Horizon Planning framework is introduced and presented in detail. 

\section{The Receding Horizon Planner}
\label{sec:rhp}

In this section, we present an extended version of the Receding Horizon Planner (RHP), first presented in \cite{rhp}, aimed at solving the Single-Task robots, Single-Robot tasks, Time-extended Assignment (ST-SR-TA) problem, as defined in \cite{gerkey2004formal}, for heterogeneous teams of agents. This problem class involves building a schedule of tasks for each agent that minimizes a given cost function and is strongly $\mc{N}\mc{P}$-hard as highlighted in \cite{brucker1999scheduling}.

The brute-force approach for solving this class of problems is to enumerate all possible schedules and choose the one with the smallest associated cost. In its simplest form, the process of generating all possible schedules is done through iteratively assigning one of the remaining tasks to each agent's schedule. As such, a set of partial schedules (i.e. schedules that do not include all tasks) will be generated to which the process is applied again. Similarly to Branch and Bound (\cite{lawler1966branch}), this process can be depicted as a tree-search where each partial schedule is associated to a node $\mc N_i$, and partial schedules generated through subsequent assignments are depicted as children of that node. However, given the set of tasks $\mc T$ and the set of agents $\mc A$, the number of possible schedules grows with $\bigO(|\mathcal{A}|^{|\mathcal{T}|})$, rendering the enumeration of all schedules intractable. A scalable alternative is the greedy approach, which solely considers the next best option given the current partial schedule. However, this approach suffers from a reduced performance of the overall system in terms of optimality.

The RHP is a task allocation scheme inspired by Model Predictive Control (MPC) which, at fixed time intervals, computes the optimal schedule for a limited number of tasks and leverages a heuristic to estimate the cost of executing the remaining tasks. Thus, the size of the optimization problem remains constant with respect to the number of tasks and can be adjusted to the computational resources available. With regard to this variable look-ahead time, the receding horizon approach is a superset of the greedy algorithms (zero look-ahead) and the full-blown optimization (infinite look-ahead). 

Similarly to MPC, the number of assignments planned by the RHP is larger than the number of assignments that are executed. This creates an overlap between the consecutive optimization cycles, which reduces the loss in optimality due to the neglected future operations. The cyclic nature of this scheme allows for the incorporation of the current system state into the optimization. This feedback loop –~as in classic control~– provides robustness against disturbances and model deviations.

\begin{figure}
    \centering
    \includegraphics[width=0.7\columnwidth]{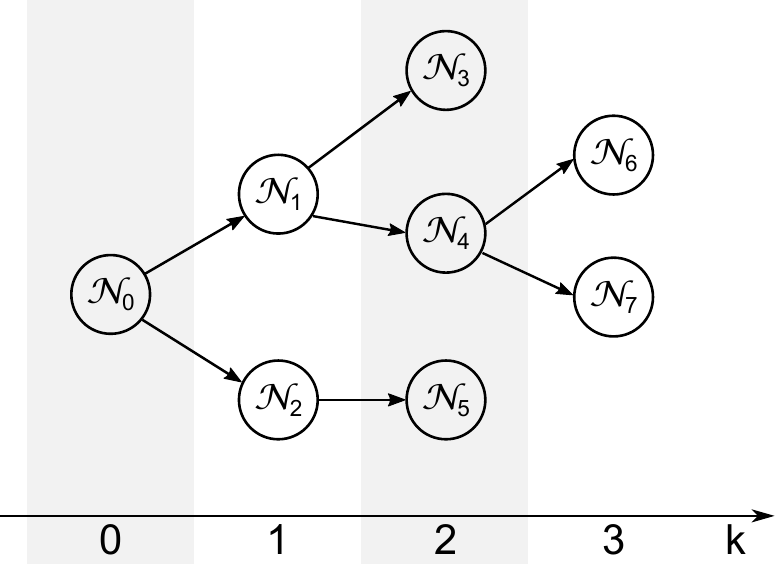}
    \caption{Example decision tree generated by the Receding Horizon Planner.}
    \label{fig:DecisionTree}
\end{figure}

As introduced in \cite{rhp}, The RHP utilizes a branch-and-bound method to generate the optimal schedule up to the desired time-horizon. As illustrated by Figure~\ref{fig:DecisionTree}, each node $\mathcal{N}_i$ in the search tree corresponds to a partial schedule and the addition of a new assignment of a task to an agent creates a new node. To enable an efficient exploration of the tree, the cost $J(\cdot)$  at node  $\mathcal{N}_i$ is decomposed into an accumulated cost value $g(\cdot)$ and a remaining cost estimation $h(\cdot)$
\begin{equation}
J(\mc N_i) = g(\mc N_i) + h(\mc N_i).
\label{eq:CostDecomposition}
\end{equation}
The accumulated cost evaluates the already assigned operations and is a measure of the consumption of resources. Since each new task assignment increases the consumption of resources, the accumulated cost increases monotonically  
\begin{equation}
    g(\mc N_j) \geq g(\mc N_i), 
    \label{eq:AccumulatedCost}
\end{equation}
where $\mc{N}_j \in Children(\mc{N}_i)$. The second summand in~\eqref{eq:CostDecomposition} is a lower-bound estimate of the remaining efforts to reach the overall goal. As each new task assignment reduces the outstanding efforts the cost of the remaining tasks must decay
\begin{equation}
    h(\mc N_j) \leq h(\mc N_i).
    \label{eq:RemainingCost}
\end{equation}
Moreover, the estimate $h$ must provide a lower bound of the true remaining cost at each step, therefore satisfying
\begin{equation}
    g(\mc N_j) - g(\mc N_i) \geq h(\mc N_i) - h(\mc N_j).
    \label{eq:LowerBoundCondition}
\end{equation}
We will show later, how such a lower bound estimation can be obtained by relaxation of the integer constraints. If $h$ is chosen such that~\eqref{eq:LowerBoundCondition} holds we can conclude that
\begin{equation}
        J(\mc N_j) \geq J(\mc N_i), 
    \label{eq:TotalCostEstimate}
\end{equation}
the cost of each node increases as the tree grows. This allows to stop the further exploration of a branch if at any time $J(\mc N_i) \geq J_{\text{opt}}$ (i.e. the cost of node $\mc N_i$ is larger than the cost of a known solution). This strategy is guaranteed to find the optimal solution on the tree.

To eliminate symmetries and thus reduce the number of nodes to be explored in a tree, only one resource (agent) is chosen for the set of offspring-nodes that are generated from any node in the tree. The agent is chosen as
\begin{equation}
a^{\ast} = \argmin_{t \in \mc T , a \in \mc A} \ (y_{ta} + T_{ta}).
\label{eq:NextAgent}
\end{equation}
The indices $t$ and $a$ tally the available tasks $\mc T$ and agents $\mc A$ respectively. $T_{ta}$ is the duration of task $t$ if performed by agent $a$ and $y_{ta}$ is the potential start time for agent $a$ on task $t$. In other words, out of all agents, the RHP chooses the one with the earliest completion-time of all tasks. Once the agent is decided upon, all potential tasks that satisfy
\begin{equation}
y_{ta}  = \min_{t \in \mc T } (y_{ta^{\ast}} + T_{ta^{\ast}}),
\label{eq:NextStates}
\end{equation}
are considered as next nodes. That is, we include all tasks that can be started, before the earliest task can be finished.  This again reduces the search space in the tree exploration without affecting the optimality of the solution.

The main contribution of this paper is to extend the range of applications of the RHP to scenarios in which multiple agents, each exhibiting different capabilities, are needed to complete a single task or vice-versa. The wildfires in our SaR scenario represent the former type, where the combined effort of multiple agents is needed to extinguish a fire. The rescue operations represent the second type, where a single agent can carry multiple survivors.

The decision space in the tree search inherently includes the various agent capabilities and teaming scenarios, if this is properly described in the set of dispatching rules for each agent. These dispatching rules describe which possible tasks an agent can do, given its current location and occupation, and to what amount the agent can contribute to the overall task, i.e. its capacity. In the context of the SaR scenario, the capacity corresponds to the mundane load capacity for water or victims of each agent.

While the implementation of these dispatching rules is straightforward, the challenge for the tree search is again limiting the search space. To avoid the exploration of all possible combinations of capacities to complete a task, we make use of the heuristic $h(\cdot)$ to guide the tree search. To achieve this, we include a high level load balancing into the heuristic, which breaks the required effort for one large task down into a set of sub-tasks that can be handled by individual agents.

The general approach is to first group all agents with equivalent capabilities into different classes and then minimize the latest finishing time among all agents under the following constraints
\begin{enumerate}[(i)]
\item \label{constraints:first} The effort for each task is distributed among the different agent classes
\item \label{constraints:second}  The efforts from~(\ref{constraints:first})  for each class are distributed among the individual member agents of this class
\end{enumerate}
The detailed application to the SaR use-case is described below.

\begin{figure*}[tb]
    \centering
    \begin{minipage}{0.33\textwidth}
    \includegraphics[width=\textwidth]{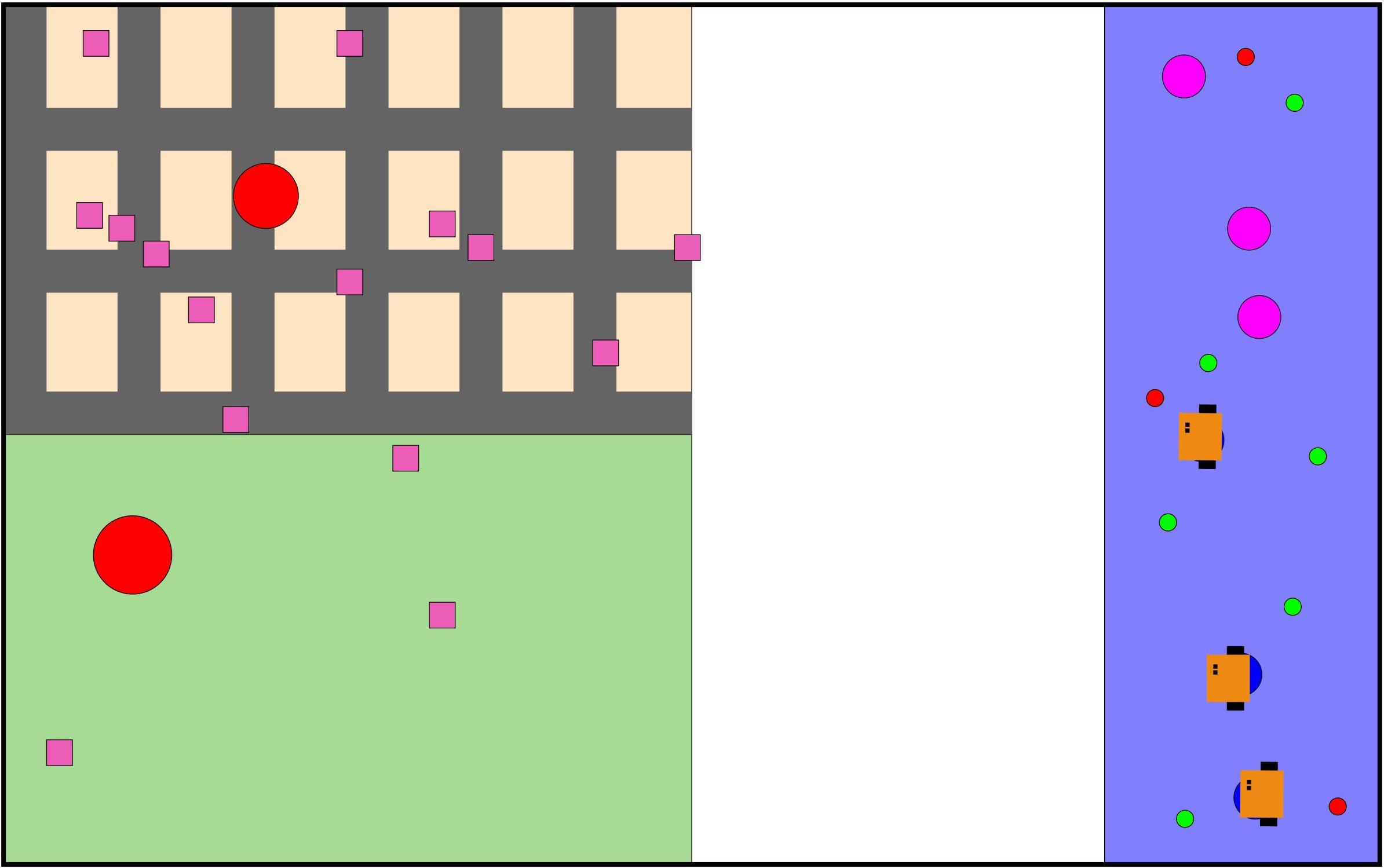}
    \end{minipage}~%
    \begin{minipage}{0.33\textwidth}
    \includegraphics[width=\textwidth]{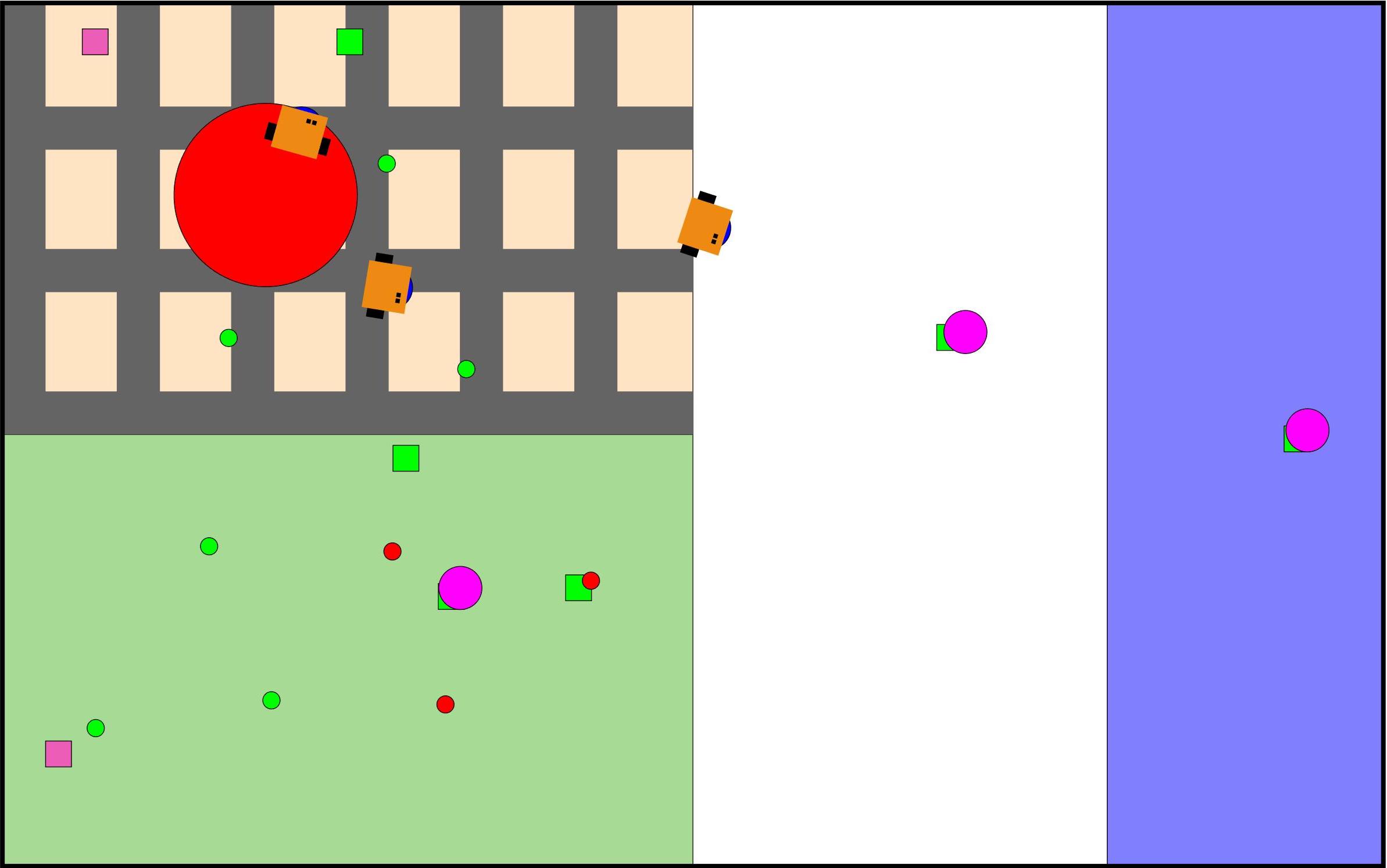}
    \end{minipage}~%
    \begin{minipage}{0.33\textwidth}
    \includegraphics[width=\textwidth]{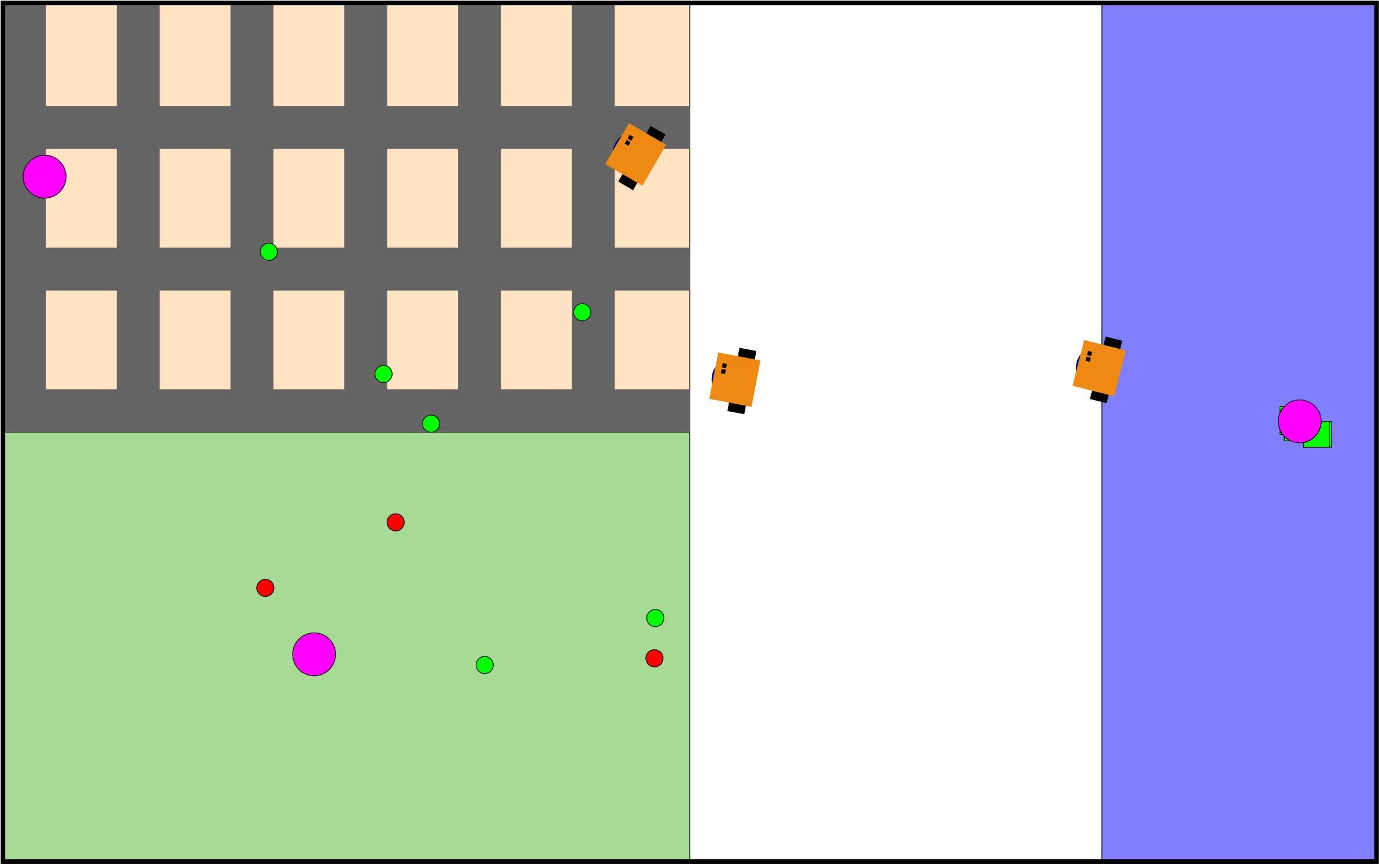}
    \end{minipage}~%
    \caption{An example scenario in the Search and Rescue Simulator. The heterogeneous team of agents is initially positioned in the starting zone (left). The orange robots denote the AGVs and the red, green and pink circles denote the ground units, drones and helicopters respectively. Hidden and identified victims are depicted as pink and green squares respectively. Fires, which can grow and spread, are depicted as red circles of radii proportional to their respective strengths. The middle and right figure depict the scenario mid and post execution respectively.}
    \label{fig:sim}
\end{figure*}

\section{Use Case: Search And Rescue}
\label{sec:useCase}

\subsection{Problem Setup}
Inspired by the RoboCup Search and Rescue Agent competition, we setup the SaR scenario as follows. A fire breaks out in a forest near a city and starts spreading. Victims are hidden within the city and forest themselves. The objective is to locate and save all victims and extinguish all fires simultaneously using a heterogeneous team of agents as quickly as possible. Note that the fires also grow and spread, therefore solely focusing on rescuing the victims is commonly a sub-optimal strategy. Moreover, the victims are initially not visible to the agents. Therefore, exploring the map is also of paramount importance.

 Each fire is represented as a circle of radius proportional to its health and requires a specific amount of water that is also proportional to its health. Upon reaching $100\%$ health, a fire then spread to a nearby territory which is illustrated through the creation of another circle in that location. In order to keep the problem setup as general as possible, we frame the scenario as a pick and place problem where agents need to repeatedly go back to the base and deliver water/victims to the fires/hospital. Moreover, we solely consider each agent's capabilities when considering it for a given task. As such, we allow for the modular addition or removal of agents from the setup as discussed in the next subsection. 

\subsection{The Heterogeneous Team}
Teams deployed for SaR are typically heterogeneous due to the variety of the tasks at hand and terrains to be navigated. Therefore, we chose to create teams composed of $4$ types of agents: Ground Units, Helicopters, Drones and Autonomous Ground Vehicles (AGVs). It is important to note that our approach is agnostic to the specific types of agents and the number existing types. In fact, the proposed approach solely considers the number of agents of each type, and each type's capabilities (as defined in \cite{gerkey2004formal}). We restrict the agent capabilities we consider to two categories. The first type deals with the mobility of the agents (e.g. what is the velocity of the agent when navigating in the forest?), and the second type considers what action the agent can perform once at the desired location (e.g. can the agent "pickup" a victim?). A tabulation of the capabilities of the $4$ types of agents is presented in Table~\ref{tab:agentCaps}. As such, each agent type's specific capabilities can be accounted for explicitly by the proposed framework in the process of generating feasible schedules. Moreover, it is worth noting that additional capabilities can be modularly added since the framework solely considers the capabilities required by each task in the process of generating new nodes. In the next subsection, we present the load-balancing Linear Program used to estimate the cost-to-go. 

\subsection{Estimation of the cost-to-go}
We formulate the cost-to-go required for the RHP as a Linear Program (LP), which added to the accumulated cost provides a lower-bound on the total cost of each node. The objective of the LP is to minimize the makespan $s$, which is the time of completion of the last task. 

In order to compute the makespan, one must be able to estimate the time taken to complete each task (e.g. rescuing a victim). This is a non-trivial problem, since the time of completion of a task by an agent is dependant upon the previous assignment of the agent. This difficulty also arises in the travelling salesman problem, where the time to travel to a given city depends on the last destination of the salesman. In order to overcome this difficulty, we assume that the distances between the victims/fires are negligible compared to their distances to the base. Therefore by lower-bounding all distances between the targets and the base, we can obtain a ``tight'' lower-bound on the amount of time a given agent takes to complete a trip to any target $t_{Type(a)}$ given the agent's velocity. 

The decision variable for the LP are
\begin{itemize}
    \item the number of assignments of task $t$ to the agents of class $c$, denoted by $n_{c,t}$
    \item the number of assignments of task $t$ to the individual agent $j$ denoted by $m_{j,t}$
    \item the total makespan denoted by $s$
\end{itemize}
With these constraints stacked into a vector
\begin{equation}
    x^{\text{T}} = [n_{c,t}\,,\,m_{j,t}\,,\, s]
\end{equation}
the LP is formulated as follows
\begin{subequations} \label{eq:heuristicLP}
\begin{align}
\min_{x} ~& \; \; [0\, \ldots\, 0\, 1] x  \label{eq:lp:a}\\ 
\subjto~~& \smashoperator{\sum_{c \in \mc C}} C_{c}^{(t)}n_{c,t} \geq R_{t} \: \forall t \in \mc T \label{eq:lp:b} \\
& y_j + \smashoperator{\sum_{t \in \mc T}} T_t^{(c)}m_{j,t} \leq s \:\: \forall j \in \mc A \label{eq:lp:c}\\ 
& \sum_{j \in \mathcal{A}}B_j^{(c)}  m_{j,t} = n_{c,t} \:\: \forall c,t \in \mc C  \otimes \mc T \label{eq:lp:d},
\end{align}
\noeqref{eq:lp:a}\noeqref{eq:lp:b}\noeqref{eq:lp:c}
\end{subequations}
where $\mc T$, $\mc C$ and $\mc A$ denote the set of all tasks, agent types and individual agents respectively. The program aims to minimize the makespan (i.e. the time of completion of the last task). Moreover, constraint \eqref{eq:lp:b} ensures that the required effort for accomplishing task $t$ denoted by $R_{t}$ is matched by the agent fleet. The effort provided by the agent fleet is computed through summing the effort provided by each agent type. The capacity of agent type $c$ for task $t$ is denoted by $C_{c}^{(t)}$.  Additionally, the makespan is computed in constraint \eqref{eq:lp:c}, where $y_j$ denotes the completion of time of agent $j$'s current schedule and the execution time of task $t$ for class $c$ is denoted by $T_{t}^{(c)}$. Lastly, constraint \eqref{eq:lp:d} ensures that the sum of contributions of each individual agent of a given type matches the total contribution of that type, where $B_{j}^{(c)}$ indicates if agent $j$ is of type $c$. In the next section, we present experimental results demonstrating the effectiveness of the SaR specialized Receding Horizon Planner presented in this section compared to a greedy scheduling approach. 

\begin{table}
\centering
 \begin{tabular}{|c||c c c c||} 
 \hline
 Types & G. Unit & Heli. & Drone & AGV \\ [0.5ex] 
 \hline\hline
 Water Cap. & 1 & 5 & 0 & 2 \\ 
 \hline
 Rescue Cap. & 1 & 4 & 0 & 4 \\
 \hline
 Move Forest & 0.1 & 0.5 & 0.40 & 0.20\\
 \hline
 Move City & 0.1 & 0.5 & 0.40 & 0.00 \\
 \hline
\end{tabular}
\vspace{3mm}
\caption{The capabilities of the $4$ types of agents: Ground Units, Helicopters, Drones and AGVs. Note for example how AGVs are unable to navigate through the forest or how the capacity for carrying water units or victims vary depending on the type of agent.}
\label{tab:agentCaps}
\end{table}


\section{Experiments}
\label{sec:experiments}
In this section, we present empirical results validating the use of a receding horizon through simulations and experiments on the Robotarium, a remotely accessible swarm accessible testbed (\cite{robotarium}). The setup of the experiments is as follows. The team is composed of 6 ground units, 3 helicopter units, 14 drones and 3 AGVs. Moreover, in each experiment $10$ initially hidden victims are randomly placed in the forest and city along with $3$ fires that were also initialized at random positions. However, the number of total fires generated in a given experiment may differ depending on the planner being used. This is due to the fact that the fires grow and spread over time and therefore require even more resources to extinguish. This is an accurate depiction of many real-world SaR scenarios and serves to emphasize that the time of completion of tasks is an important measure in such scenarios.  

 The path-action planning algorithm details the execution of the generated schedules and is implemented as follows. First, it generates each agent's path to its corresponding task, then ensures that the agent takes the corresponding required action for the task if feasible. For example, once an agent tasked with rescuing a victim reaches its location, it will take the "pick-up" action if and only if the agent's maximum capacity for the number of victims is not already reached. Since path planning is not the focus of this work, we assume all agents except the AGVs possess single-integrator dynamics and use  proportional controllers to guide them to their targets. However, since the AGVs are presented as robots (GRITSBot X) in the Robotarium experiments and can indeed collide, we implemented multi-agent A* to generate way-points the agents can follow to their targets. Moreover, since we do not explicitly check for collision-avoidance in the trajectories between way-points, we also utilize Control Barrier Functions (CBFs) to instantaneously ensure collision-avoidance at all times as described in \cite{AmesBarriers, ames2014}. This is achieved through solving a Quadratic Program at each point in time that generates a minimally altered trajectory for the agents relative to their nominal trajectory to ensure collision-avoidance. 
 
A run of $20$ simulated experiments with randomized initial conditions were run to compare several planning depths of the RHP and a greedy scheduling approach. The greedy scheduling scheme used for bench-marking the proposed RHP is a one-step look-ahead planning approach. Specifically, at each scheduling iteration, each idle agent is assigned to the task that it is closest to. The mean, median and variance of the makespans of each of the scheduling approaches over all experiments are presented in Table~\ref{tab:results}. As shown in the table, the mean makespan decreases significantly when the RHP is used. However, the rate of improvement decreases at higher planning depths, which highlights the trade-off between computing time and solution quality.

\begin{figure}[t]
\centering
 \includegraphics[width=0.40\textwidth]{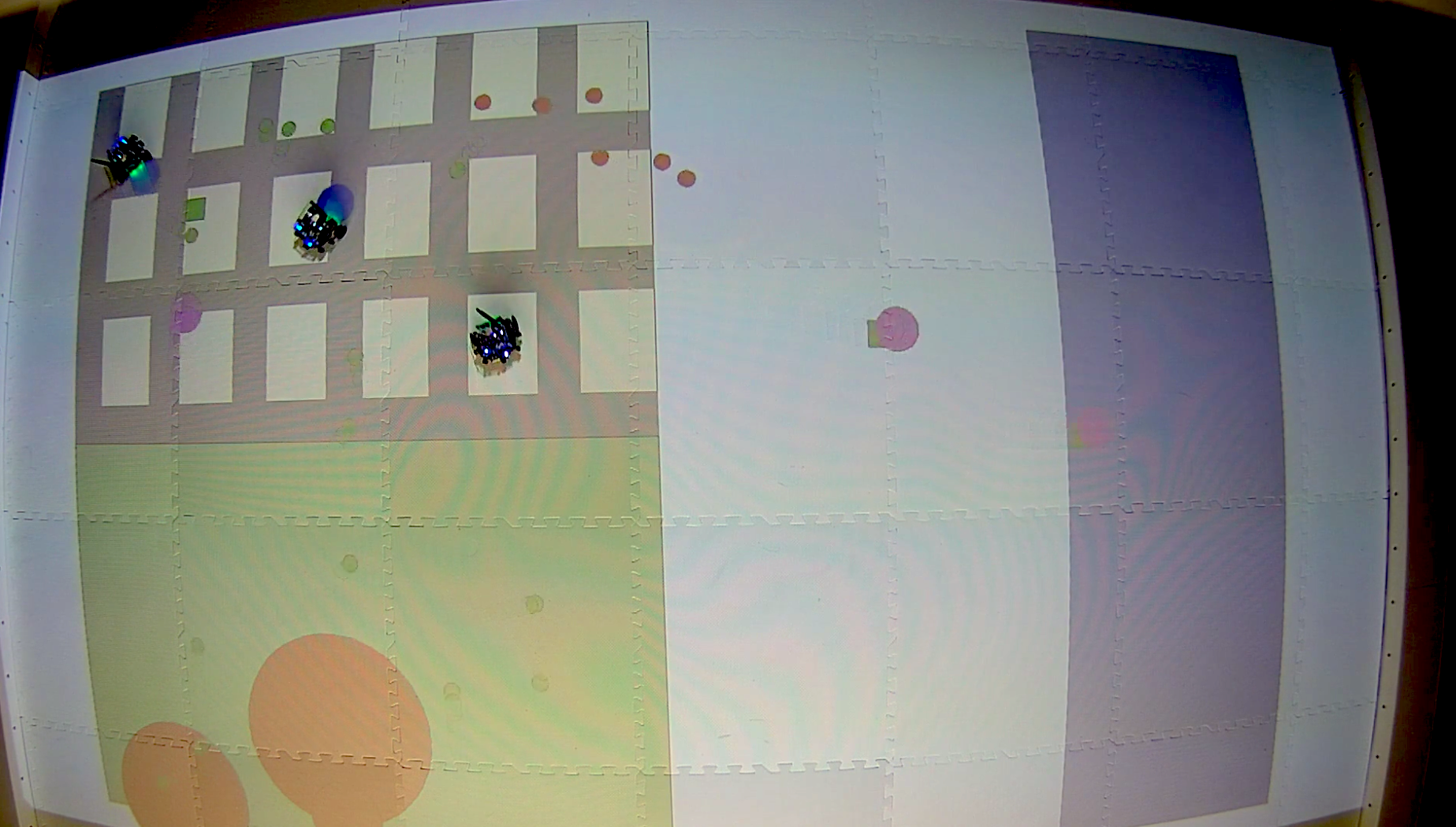}
 \caption{Search and Rescue experiment on the Robotarium using 3 GRITSBot X as the AGVs. The description of all entities are as in Figure ~\ref{fig:sim}.}
 \label{fig:robotariumExp}
\end{figure}

Moreover, to demonstrate the applicability of the RHP onto real systems, $10$ experiments were conducted on the Robotarium testbed comparing the RHP using a planning depth of $10$ to the greedy scheduler. The Robotarium's robots (GRITSBot X) were used instead of the three simulated AGVs as depicted in Figure~\ref{fig:robotariumExp}. To obtain the desired frequency of operation on the Robotarium ($\sim100$~Hz), a separate computing node was used to run the scheduling algorithms, and the schedules were transmitted to the agents using a publisher-subscriber protocol. The results of the experiments are presented in Table~\ref{tab:resultsR}. Indeed, the use of the RHP reduces the makespan, thus validating the applicability of the proposed scheduling approach.

\begin{table}
\centering
 \begin{tabular}{|c||c c c||} 
 \hline
 Types & Mean & Median & $\sigma$ \\ [0.5ex] 
 \hline\hline
 Greedy & 78.78 & 53.03 & 10.63  \\
 \hline
 RHP10 & 42.54 & 40.34  & 9.54  \\ 
 \hline
  RHP15 & 40.09 & 38.56 &  7.95 \\ 
  \hline
  RHP20 & 39.87 & 36.84 &  9.39 \\ 
 \hline
\end{tabular}
\vspace{3mm}
\caption{Results over $20$ simulations comparing the mean, median and standard deviation of the makespans of the Receding Horizon Planner with different planning depths and greedy scheduling scheme.}
\label{tab:results}
\end{table}

\begin{table}
\centering
 \begin{tabular}{|c||c c c||} 
 \hline
 Types & Mean & Median & $\sigma$ \\ [0.5ex] 
 \hline\hline
 Greedy & 46.70  &  45.06  & 9.64 \\
 \hline
 RHP10 & 39.00 &  36.72 & 9.09 \\ 
 \hline
\end{tabular}
\vspace{3mm}
\caption{Results over $10$ Robotarium experiments comparing the mean, median and variance of the makespans of the Receding Horizon Planner with planning depth $10$ and the greedy scheduling scheme.}
\label{tab:resultsR}
\end{table}

\section{Conclusion}
\label{sec:conclusion}
This paper introduces a task allocation framework capable of scheduling tasks for heterogeneous teams of agents in a manner that is tractable and robust to changes in the environment and in the agent fleet. This was achieved through repeatedly scheduling solely up to a fixed horizon and leveraging a load-balancing Linear Program for the estimation of the cost to go. Moreover, a simulation framework for an abstracted Search and Rescue scenario inspired by the RoboCup Search and Rescue Agent Simulation competition was presented along with a specialized formulation of the Receding Horizon Planning approach. Experimental results showcase the efficacy of the proposed scheduling method in extensive multi-agent simulations and experiments on the Robotarium.

\bibliography{ifacconf}             

\end{document}